\documentclass[manuscript]{emulateapj}

\usepackage{apjfonts}

\slugcomment{to appeear in the Astrophysical Journal}

\shorttitle{Super-Eddington Luminosity Model of IUE Novae}
\shortauthors{Kato \& Hachisu}

\begin{document}


\title{MODELING OF THE SUPER-EDDINGTON PHASE FOR CLASSICAL NOVAE: FIVE IUE 
NOVAE}

\author{Mariko Kato}
\affil{Department of Astronomy, Keio University, 
Hiyoshi 4-1-1, Kouhoku-ku, Yokohama 223-8521, Japan:}
\email{mariko@educ.cc.keio.ac.jp}

\and

\author{Izumi Hachisu}
\affil{Department of Earth Science and Astronomy, 
College of Arts and Sciences,
University of Tokyo, Komaba 3-8-1, Meguro-ku, Tokyo 153-8902, Japan;}
\email{hachisu@chianti.c.u-tokyo.ac.jp}




\begin{abstract}
We present a light curve model for the super-Eddington luminosity phase 
of five classical novae observed with {\it IUE}. Optical and UV light curves 
are calculated based on the optically thick wind theory with 
a reduced effective opacity for a porous atmosphere. 
Fitting a model light curve with the UV 1455 \AA~ light curve,  
we determine the white dwarf mass and distance to be 
$(1.3~M_{\odot}$, 4.4 kpc) for \object{V693 CrA}, 
$(1.05~M_{\odot}$, 1.8 kpc) for \object{V1974 Cyg}, 
$(0.95~M_{\odot}$, 4.1 kpc) for \object{V1668 Cyg}, 
$(1.0~M_{\odot}$, 2.1 kpc) for \object{V351 Pup}, 
and $(1.0~M_{\odot}$, 4.3 kpc) for \object{OS And}.
\end{abstract}





\section{INTRODUCTION}

The super-Eddington luminosity is one of the long standing problems in 
the theoretical study of classical novae 
\citep[e.g.,][for recent summary]{fri04}.
Super-Eddington phases last more than a few to several days, 
and their peak luminosities often exceed the Eddington limit 
by a factor of a few to several
\citep[e.g.,][and references therein]{del95}.


It is difficult to reproduce such a super-Eddington luminosity in 
evolutional calculations of nova outbursts. 
Dynamical calculations show that the super-Eddington phase 
appears only in a very short time, or does not appear at all,
apart from numerical difficulties that often prevent accurate 
calculation of the photospheric luminosity and visual magnitude 
\citep[e.g.,][]{spa78, pri78, nar80, sta85, sta86, pol95}.

Recently, \citet{sha01b,sha02} presented an idea on the mechanism of 
the super-Eddington luminosity. Shortly after hydrogen ignites on a white 
dwarf, the envelope becomes unstable to develop a porous  
structure in which the effective opacity becomes much smaller 
than the normal opacity for uniform medium.
Corresponding to the reduced effective opacity, 
the effective Eddington luminosity becomes larger.
Therefore, the diffusive energy flux can exceed the Eddington 
value for uniform medium, even though it does not exceed the 
effective Eddington luminosity.

Based on this idea, \citet{kat05} presented a light curve model for the 
super-Eddington phase of classical novae. They assumed a reduced opacity in early 
phases of nova outbursts and reproduced the optical light curve of V1974 Cyg. 
This is the first theoretical model for the super-Eddington light curves.
In the present paper, we apply the same method to other classical novae 
to examine whether or not this idea is applicable to the different speed 
class of novae. 

The International Ultraviolet Explorer  ({\it IUE}) satellite
observed a number of nova outbursts 
\citep[e.g.,][]{cas79,cas02,cas04a,cas05,sti81}.
\citet{cas02} presented  1455 \AA~ continuum
light curves for twelve novae and showed that the duration of the UV 
outburst is a good indicator of the speed class of novae, i.e., a faster nova
shows a shorter duration of the UV outburst.
This 1455 \AA~ light curve is an important clue in modeling the 
super-Eddington phase \citep{kat05} and also a useful tool
in estimating the white dwarf mass and distance to the star  
\citep{hac05k,hac06}.
Here, we make a model of the super-Eddington phase
for five Galactic novae, V693 CrA, V1974 Cyg, V1668 Cyg, 
V351 Pup, and OS And, because  their 1455 \AA~ light curves
are available from the beginning of the super-Eddington phase to 
the UV decay.

Section \ref{lightcurve_model} gives a brief description of our
method for UV light-curve fittings. The light curve analyzes 
for individual objects are shown in \S \ref{V693 CrA}--\ref{osand}.
In \S \ref{discussion} we summarize our results.

\section{LIGHT CURVE MODEL} \label{lightcurve_model}

\subsection{Optically Thick  Wind Model}
 \label{opticall_thick_wind_model}


After a thermonuclear runaway sets in on an accreting white dwarf (WD), 
the photosphere greatly expands to $R_{\rm ph} \gtrsim 100 ~R_{\odot}$.
The optical luminosity reaches a maximum value, which often 
exceeds the Eddington limit. After that, the photosphere moves
inward whereas the envelope matter goes outward. 
The wind mass-loss begins in the very early phase of the outburst and 
continues until the photospheric temperature rises to $\log T_{\rm ph} 
\sim 5.2-5.6$. The envelope mass decreases owing to the 
wind and nuclear burning  \citep{kat94h}. 

The decay phase of novae can be well represented with a sequence 
of steady state solutions as described by \citet{kat94h}. 
We have solved a set of equations, i.e., the equations
of motion, mass continuity, radiative diffusion, and conservation of energy,
from the bottom of the hydrogen-rich envelope through the photosphere.
The winds are accelerated deep inside the photosphere, so 
they are called ``optically thick winds.''

\subsection{The Reduced Opacity}

We assume that the opacity is effectively 
reduced by a factor $s$,
\begin{equation}
\kappa_{\rm eff} = \kappa/s,
\label{opac_equation}
\end{equation}
in the super-Eddington phase, where $\kappa$ is the OPAL opacity \citep{igl96} 
and $s$ is the opacity
reduction factor that represents the reduced ratio of the effective opacity 
in a porous envelope. 


\citet{kat05} assumed that $s$ is a function of the temperature
and time, i.e.,  $s$ is unity in the outer region of the envelope
($\log T < 4.7$), but takes a certain constant value $s_0~( > 1)$
at the inner region $\log T > 5.0$, 
and it changes linearly between these values. 
Here $s_0~$ is a function of time that has the maximum value at the optical 
peak and then gradually decreases to unity. Choosing an appropriate 
$s(T,t)$,  Kato \& Hachisu reproduced the light curve of V1974 
Cyg in the super-Eddington phase. Once the temperature dependence of 
$s$ is given, $s_0$ is uniquely determined by 
fitting with both the optical and UV light curves. 
In the present paper, we first adopt the same temperature dependence of $s$ 
as in \citet{kat05}. We call it Model 1.

We adopt another type of function for $s$.  We call it
Model 2, in which $s$ changes in a more inner region of the 
envelope, i.e.,  $s=1$ at  $\log T < 5.25$  but  $s=s_0$ at $\log T > 5.45$ 
and changes linearly between these values.

The function of $s$ should be closely linked with radiation
instabilities against a porous structure of the atmosphere.
However, we do not know how and when the porous structure develops  
in a nova envelope and how much the opacity is reduced. Therefore, in the 
present paper, we assume the two functions of $s$, i.e., Model 1 and Model 2.
Model 1 corresponds to the case that the porous 
structure develops from the bottom 
of the envelope to a lower temperature region beyond 
the peak of the OPAL opacity at $\log T \sim 5.2$.
Therefore, the peak value of the opacity is reduced by a factor of $s_0$. 
Model 2 corresponds to the case that the porous structure does not 
extend to the opacity peak at $\log T \sim 5.2$.

\subsection{Optical Light Curve} \label{opticalcurve}

In V1500 Cyg, which is one of the brightest novae, 
the temporal evolution of the spectrum and the fluxes are well understood 
as blackbody emission during the first three days and 
as free-free emission after that 
\citep{gal76,enn77,due79}. 
In the modeling of the super-Eddington phase we divide the optical 
light curve into three phases \citep{kat05}.
The first is the super-Eddington phase, in which we simply 
assume that photons are emitted at the photosphere as a blackbody 
with a photospheric temperature of $T_{\rm ph}$.
In the next phase, the optical flux is 
dominated by free-free emission of the optically thin ejecta 
outside the photosphere.
The flux of free-free emission can be roughly estimated as
\begin{equation}
F_\lambda \propto \int N_e N_i d V
\propto \int_{R_{\rm ph}}^\infty {\dot M_{\rm wind}^2 \over {r^4v^2}} r^2 dr
\propto {\dot M_{\rm wind}^2 \over {R_{\rm ph}v_{\rm ph}^2}},
\label{free-free-wind}
\end{equation}
where $F_\lambda$ is the flux at the
wavelength $\lambda$, $N_e$ and $N_i$ are the number densities of
electrons and ions, $V$ is the volume of the ejecta,
$\dot M_{\rm wind}$ is the wind mass-loss rate,
and  $v_{\rm ph}$ is the velocity at the photosphere. Here, we use
the relation of $\rho_{\rm wind} = \dot M_{\rm wind}/ 4 \pi r^2
v_{\rm wind}$, and $\rho_{\rm wind}$ and $v_{\rm wind}$ are
the density and velocity of the wind, respectively.
We substitute $\dot M_{\rm wind}$, 
$R_{\rm ph}$ and $v_{\rm ph}$ from our best fit model.
We cannot uniquely specify the proportional constant in equation 
 (\ref{free-free-wind})
because radiative transfer is not calculated
outside the photosphere.  Instead,
we choose the constant to fit the light curve \citep{hac05k,hac06}.

When the nova enters a nebular phase, strong emission lines such as 
[\ion{O}{3}] dominantly contribute to the visual light curve. 
Then the visual light curve gradually deviates from our free-free light 
curve of equation (\ref{free-free-wind}). This is the third phase.

\subsection{UV 1455 \AA~ Light Curve} \label{uv_flux}

After the optical maximum, the photospheric radius of the envelope 
gradually decreases while the photospheric temperature ($T_{\rm ph}$) 
increases with time.
As the temperature increases, the main emitting wavelength of 
radiation shifts from optical to UV.
The UV 1455 \AA~flux reaches a maximum at $\log T_{\rm ph} \sim 4.4$.
After the UV flux decays, the supersoft X-ray flux finally increases.


\begin{figure}
\epsscale{1.15}
\plotone{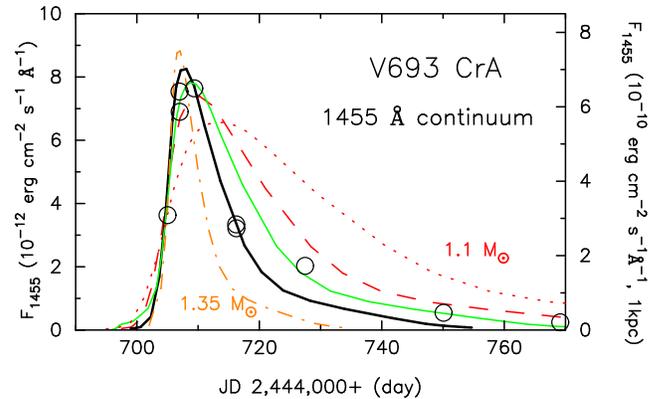}
\caption{
UV light curve-fitting for V693 CrA. Calculated 1455 \AA~ continuum fluxes
are plotted against time for the WD models of  $M_{\rm WD}=$ 1.1 $M_{\odot}$(dotted line), 
1.2 $M_{\odot}$ (dashed line), 1.25 $M_{\odot}$ (thin solid line), 1.3 $M_{\odot}$ (thick solid line), and 1.35 
$ ~M_\odot$ (dash-dotted line). The chemical composition of the envelope 
is assumed to be $X=0.35,~Y=0.33,~X_{\rm CNO}=0.2$,
$X_{\rm Ne}=0.1$, and $Z=0.02$.  
The theoretical flux $F_{1455}$ is calculated for an arbitrarily assumed distance 
of 1.0 kpc and no absorption (scale in the right-hand-side).
Open circles denote the 1455 \AA~ continuum flux taken from \citet{cas02}.
\label{v693craUV}
}
\end{figure}

Figure \ref{v693craUV} shows the theoretical 1455 \AA~ light-curves
of various WD masses with an envelope chemical composition
 of  $X=0.35,~Y=0.33,~X_{\rm CNO}=0.2, ~X_{\rm Ne}=0.1,$ and $Z=0.02$. 
The evolutional timescale depends strongly on the WD mass.
More massive WDs evolve faster than less massive WDs.
The evolution speed also depends on the chemical composition 
because enrichment of heavy elements drives more massive winds 
through the opacity enhancement, which accelerate nova evolutions.
Therefore, the duration of 1455 \AA~ burst depends on the 
WD mass and chemical composition of the envelope.

\begin{figure}
\epsscale{1.15}
\plotone{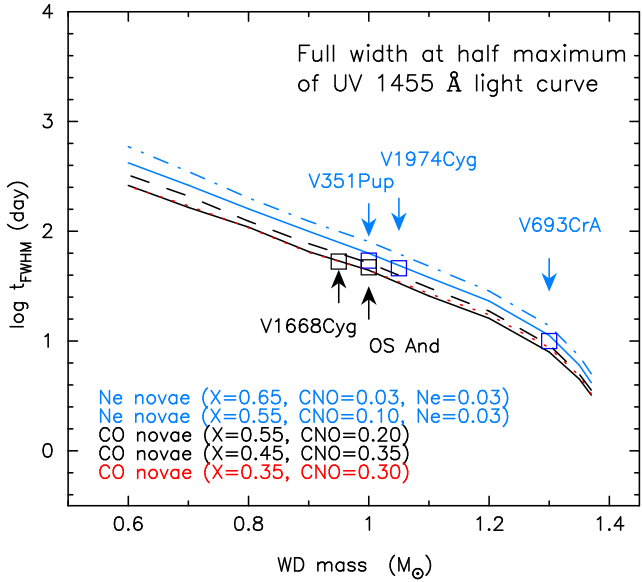}
\caption{
Full width at the half maximum (FWHM) for the 1455 \AA~ light curve vs.  
the WD mass. Dash-dotted line:
neon novae ($X=0.65,~ Y=0.27,~ X_{\rm CNO}=0.03,~ X_{\rm Ne}=0.03$). 
Solid line: neon novae ($X=0.55,~ Y=0.30,~ X_{\rm CNO}=0.10,~ X_{\rm Ne}=0.03$).
Dashed line: CO novae ($X=0.55,~ Y=0.23,~ X_{\rm CNO}=0.20$).
Solid line: CO novae ($X=0.45,~ Y=0.18,~ X_{\rm CNO}=0.35$).
Dotted line: CO novae ($X=0.35,~ Y=0.33,~ X_{\rm CNO}=0.30$), which is 
almost overlapped with the lower solid line. We assume $Z=0.02$ for all the models.
Arrows indicate the WD mass for each object. 
\label{UVwidth}
}
\end{figure}

Figure \ref{UVwidth} depicts the dependence of such duration of the 1455 \AA~ 
outburst for various WD masses and chemical compositions. Here, 
the duration is defined by the  
full width at the half maximum of the 1455 \AA~ light curve \citep{hac06}. 
Once the chemical composition is determined, we can estimate the 
WD mass from this figure. 


\citet{hac06} fitted theoretical light curves with the 1455 \AA~ 
observation and determined the WD masses for V1668 Cyg
and V1974 Cyg. They find that their WD masses show 
good agreement in the light-curve fittings of optical, infrared,  
and X-ray for the entire period of the outburst.
In this sense, the 1455 \AA~ light curve is a good indicator
of the WD mass. In the present paper, we estimate the WD mass from  
the 1455 \AA~ light-curve fitting and use it in modeling of 
the super-Eddington phase.






\begin{deluxetable*}{llllllll}
\tablecaption{Model Parameters \label{model_parameters}}
\tablewidth{0pt}
\tablehead{
\colhead{object } & \colhead{V693 CrA}&\colhead{ }& \colhead{V1974 Cyg}&\colhead{ } & \colhead{V1668 Cyg}
& \colhead{V351 Pup}& \colhead{OS And}
\\ \colhead{outburst year} & \colhead{1981}&\colhead{ }& \colhead{1992}&\colhead{ }& \colhead{1978}&\colhead{1991}& \colhead{1986}
}
\startdata
opacity Model (1 or 2) & 1&2&1& 2& 2&2& 2\\
$M_{\rm WD}$~(M$_{\odot}$)&1.3&$\leftarrow$&1.05&$\leftarrow$&0.95&1.0&1.0  \\
$ X$     & 0.35&$\leftarrow$ &0.46 &$\leftarrow$ &0.45& 0.35 &0.45  \\
$Y $     & 0.33& $\leftarrow$&0.32 &$\leftarrow$ &0.18& 0.23 &0.18  \\
$X_{\rm CNO}$      & 0.2&$\leftarrow$& 0.15&$\leftarrow$ &0.35& 0.3  &0.35  \\
$X_{\rm Ne} $     & 0.1  &$\leftarrow$&  0.05&$\leftarrow$ & 0&0.1 & 0  \\
$Z $      & 0.02 &$\leftarrow$ &  0.02 &$\leftarrow$& 0.02 & 0.02 &0.02  \\
$E(B-V)^{\rm a}$  &0.2 &$\leftarrow$&0.32 &$\leftarrow$&$0.40$  
         &$0.72$  &$0.25$     \\
$s_0$ at peak$^{\rm b}$     & 2.2 & 2.7 &5.0& 7.2 & 9.0 & 5.5 &6.0   \\
Distance (kpc)             & 4.4  &4.4 &1.8&1.8 &4.1 & 2.1 & 4.3 \\
UV FWHM$^{\rm c}$ (days)            &10 &10& 46& 46 & 53 & 54 & 47  \\
$L_{\rm max}~(10^{38}$ erg~${\rm s}^{-1})$ &2.68&2.65&4.4&4.2&4.0&3.3&3.2 \\
$M_{V, \rm max}$~(mag)&$-7.30$& $-7.30$ &$-7.75$&$-7.73$&$-7.73$&$-7.43$&$-7.56$ \\
$m_{V, \rm max}$~(mag)$^{\rm d}$& 6.5 & 6.5 & 4.6 &  4.6 &  6.6   & 6.4  &6.4  \\
excess of super-Edd (mag)           &  0.84&  0.84 &1.7 & 1.7 & 1.9 & 1.5 & 1.6 \\
duration of super-Edd (days)          &   6   & 6    & 18$^{\rm e}$ & 16$^{\rm e}$  & 16  & 9  & 12 \\
$t_3$ time (days)        & 13 & 13 &34$^{\rm e}$ & 32$^{\rm e}$ & 25 & 23 &25 \\
$\Delta M_{\rm eject}$~($10^{-5} M_{\odot}$) & 2.0 &2.7&3.9&4.8&5.8 & 3.1 &3.9  \\
\enddata
\tablenotetext{a}{\citet{van97} for V693CrA; \citet{cho97} for V1974 Cyg;
 \citet{hac06} for V1668 Cyg; \citet{sai96} for V351 Pup; \citet{sch97} for OS And 
}
\tablenotetext{b}{opacity reduction factor in equation 
(\ref{opac_equation}) at the optical maximum}
\tablenotetext{c}{calculated from our theoretical UV 1455 \AA~light curve}
\tablenotetext{d}{apparent magnitude corresponds to $M_{V}^{\rm peak}$ in the light
curve fitting}
\tablenotetext{e}{estimated from the part of free-free light curve}
\end{deluxetable*}

\section{V693 CrA (NOVA CORONAE AUSTRALIS 1981)} \label{V693 CrA}

Nova \object{V693 CrA} was discovered by Honda \citep{koz81,cal82} on 
1981 April 2 near maximum at an apparent magnitude of $6.5$.
The discovery magnitude was once reported to be 7.0 but was later 
corrected to be 6.5 \citep[Kozai, private communication in][]{cal82}.
\citet{sio86} suggested a white dwarf as massive as those in O-Ne-Mg
novae from the broad emission line widths,
high ejection velocities, large mass-loss rates,
and the presence of strong neon lines \citep{van97}. 

The abundance of V693 CrA  was estimated from the {\it IUE} spectra
to be $X=0.38,~Y=0.20$, $X_{\rm CNO}=0.15$, and 
$X_{\rm Ne}=0.26$ \citep{van97}, 
$X=0.26,~Y=0.30,$ $X_{\rm CNO}=0.33$, and $X_{\rm Ne}=0.20$ 
\citep[][taken from Table 6 in Valandingham et al. 1997
]{wil85}, or
$X=0.16,~Y=0.18,$ $X_{\rm CNO}=0.36$, and $X_{\rm Ne}=0.27$  
\citep{and94}. Considering these scattered values, we adopt 
$X=0.35,~Y=0.23,~X_{\rm CNO}=0.22$, $X_{\rm Ne}=0.10$ and $Z=0.02$ 
in our model calculation.
The theoretical light curve hardly changes if we increase the
neon abundance from 0.1  to 0.2 and decrease the helium from 0.23 to 0.13.
 This is because the exchange of neon with helium does not affect either the 
hydrogen burning rate or the opacity.

\subsection{UV Light Curve and Distance}
 
    Figure \ref{v693craUV} depicts the {\it IUE} continuum UV fluxes 
obtained by \citet{cas02} for the $1455$~\AA~ band with a
$\Delta \lambda=20$~\AA~ width (centered on $\lambda=1455$~\AA).
The corresponding theoretical light curves are also plotted 
for five WD masses of 1.1, 1.2, 1.25, 1.3, and 1.35 $M_{\odot}$.
Here, we assume the OPAL opacity (i.e., $s=1$ throughout the envelope). 
For more massive WDs the evolution is faster and the UV flux decays 
more quickly. 
Both the 1.25 and 1.3 $M_{\odot}$ WDs are consistent 
with the observation. We may 
exclude $M_{\rm WD} < 1.2 M_{\odot}$ and $M_{\rm WD} > 1.35 M_{\odot}$. 
In the present paper,we adopt the 1.3 $M_{\odot}$ model for 
later calculation.

 \citet{van97} obtained the reddening of V693 CrA to be $E(B-V)=0.2 \pm 0.1$
mainly from the comparison with nova LMC 1990 No. 1 and also the comparison 
with the reddening of globular clusters within 10 degrees from the nova. 
They also suggested that the reddening is small because no interstellar 
absorption feature was seen in their spectrum. In the present paper, 
we assume $E(B-V)=0.2$.

Using this reddening, we estimate the distance to the star.  
The absorption at $\lambda = 1455$ \AA~ is calculated to be
$A_\lambda = 8.3 ~E(B-V) =1.66 $ \citep[e.g.,][]{sea79}. The observed 
peak flux is  $7.64\times 10^{-12}$
  erg~cm$^{-2}$~s$^{-1}$\AA$^{-1}$ whereas the theoretical 
peak value is $7.02 \times 10^{-10}$ erg~cm$^{-2}$~s$^{-1}$\AA$^{-1}$ 
for the 1.3 $M_{\odot}$ star for a 
distance of 1 kpc. Therefore, the distance is calculated to be 
$D=\sqrt{7.02\times 10^{-10}/7.64
\times 10^{-12}/10^{(1.66/2.5)}}
=4.4$~kpc.

In previous works, the distance to V693 CrA has been estimated to be as 
large as 8-12 kpc \citep{cal81,cal82,bro81,bro82}, mainly because 
they assumed large absolute magnitudes of $M_{\rm V}=-8.75$ to $-10$ 
from the absolute magnitude-rate of decline relations.
This relation, however, is not very accurate for a single nova \citep{van97}
and may overestimate the absolute magnitude. We will see that 
the peak magnitude is as faint as $M_V= -7.30$ in both Models 1 and 2 
for the obtained distance of 4.4 kpc.  We summarize our fitting 
results in Table \ref{model_parameters}.


\begin{figure}
\epsscale{1.15}
\plotone{f3.epsi}
\caption{
Light-curve fitting for V693 CrA 1981.
Thick solid lines: calculated V light curves for the blackbody assumption.
Thin solid lines: calculated 1455 \AA~ light curves.
The WD mass is assumed to be $ 1.3~M_\odot$ with the envelope chemical 
composition of  $X=0.35, ~Y=0.33, 
~X_{\rm CNO}=0.2,~X_{\rm Ne}=0.1$ and $Z=0.02$.  
Optical data are taken from AAVSO (crosses), \citet[][squares]{cal81},
{\it IUE} VFES \citep[][asterisks]{cas04a}, and IAUC 3591, 3590, 3594, and 
3604 (small open circles and arrows for upper limit 
observation before the outburst). The maximum magnitude on JD 2,444,697 was 
corrected from 7.0 to 6.5  \citep{cal82}. The 
 1455 \AA~data are the same as in Fig. \ref{v693craUV}. 
The opacity reduction factor $s_0$ is plotted by a thin solid curve
in a linear scale between 1.0 (bottom) and 2.2 (at the peak)
in the upper panel (Model 1), and 
1.0 (at the bottom) and 2.7 (at the peak) in the lower panel (Model 2).
The distance of 4.4 kpc is assumed, 
which is estimated from the 1455 \AA~ light-curve fitting.
\label{v693light}
}
\end{figure}

\subsection{Optical Light Curve}

Figure \ref{v693light} shows the theoretical light curves as well as the 
observational data in optical. Because of few data points  
around the maximum and of scattered data after that, it is difficult to 
identify a shape of the optical light curve. Our best fit 
model for each opacity reduction factor shows good agreement with both 
the UV and optical light curves.
These two models have similar properties as shown in Table 1. 

In Figure \ref{v693light}
the thick solid curve denotes the visual light curve for 
the blackbody photosphere. The super-Eddington phase lasts
6 days in both models.
Free-free emission gradually becomes dominant as the photospheric 
temperature rises to $\log T > 4.0$ and
our theoretical curve for blackbody emission 
deviates from the observation. Here, we do not plot the light curve of 
free-free emission phase, because we cannot exactly determine 
the constant in equation (\ref{free-free-wind})
for such large scattering data.

In the case of V693 CrA the super-Eddington phase almost  
ends before the 1455 \AA~ flux rises. Therefore, our estimates of 
the WD mass and distance, which are determined from the UV flux
fitting, are probably independent 
of our modeling of the super-Eddington phase.

\placetable{model_parameters}

\section{V1974 Cyg (NOVA CYGNI 1992)} \label{v1974cyg}

\begin{figure}
\epsscale{1.15}
\plotone{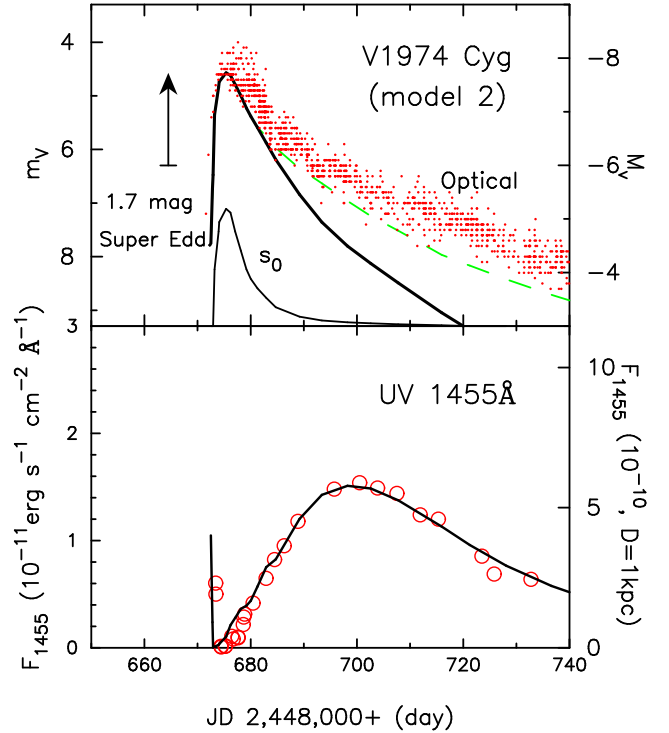}
\caption{
Light-curve fitting for V1974 Cyg 1992 for Model 2. (a) Upper panel. 
Thick solid line: $V$-magnitude from the blackbody photosphere.
Dashed line: $V$-magnitude from the free-free emission 
calculated from equation (\ref{free-free-wind}).
Thin solid line: the opacity reduction factor $s_0$ 
in the linear scale between 1.0 (at the bottom) and 7.2 (at the peak). 
Optical data are taken from AAVSO (dots). (b) Lower panel. Open circles: The
 1455 \AA~ data taken from \citet{cas04b}. 
In the upper panel the distance
of 1.8 kpc is assumed, which is obtained from fitting in the lower panel.
\label{v1974cyg2}
}
\end{figure}

\citet{kat05} presented a light curve model of the super-Eddington phase
for a 1.05 $M_{\odot}$ WD with a chemical composition of $X=0.46$,
$X_{\rm CNO}=0.15$, $X_{\rm Ne}=0.05$ and $Z=0.02$.
This is the prototype of Model 1.  
Here, we have calculated light curves for Model 2
with the same other parameters as in Kato \& Hachisu.
Our best fit model in Figure \ref{v1974cyg2} is very similar to
 that for Model 1.  This means that we cannot 
determine which opacity reduction factor $s$ is preferable 
for V1974 Cyg from the light-curve fitting.


The distance is calculated from the UV light-curve fitting.
We obtained 1.8 kpc with an  extinction of $E(B-V)= 0.32$ \citep{cho97}, 
the same value as in  Model 1.
Our distance is slightly larger than the estimate 
of 1.7~kpc by \citet{hac05k} obtained from the UV light-curve fitting using the 
normal opacity ($s \equiv 1$) models,  
because the luminosity at the UV peak is still super-Eddington 
($s_0=1.15$ in Model 1 and $s_0=1.2$ in Model 2 at the UV peak),   
and the UV flux is $15-20 \%$ larger than that in the normal opacity models.
Our distance is consistent with those discussed in  
\citet{cho97}, a most probable value of 1.8 kpc.

\section{V1668 CYG (NOVA CYGNI 1978)} \label{v1668cyg}

V1668 Cyg was discovered on 1978 September 10.24 UT 
\citep{mor78}, two days before its optical maximum
of $m_{V, {\rm max}} = 6.04$. This object was also well observed with 
{\it IUE} satellite. \citet{hac06} presented a light curve model  
from shortly after the optical peak  until the end of the 
outburst. Their best fit model is a WD mass of $0.95~M_{\odot}$ with a 
chemical composition of $X= 0.45$, $X_{\rm CNO}= 0.35$,
and  $Z= 0.02$. 
In the present paper, we adopt their parameters and reproduce 
the super-Eddington phase. 


\begin{figure}
\epsscale{1.15}
\plotone{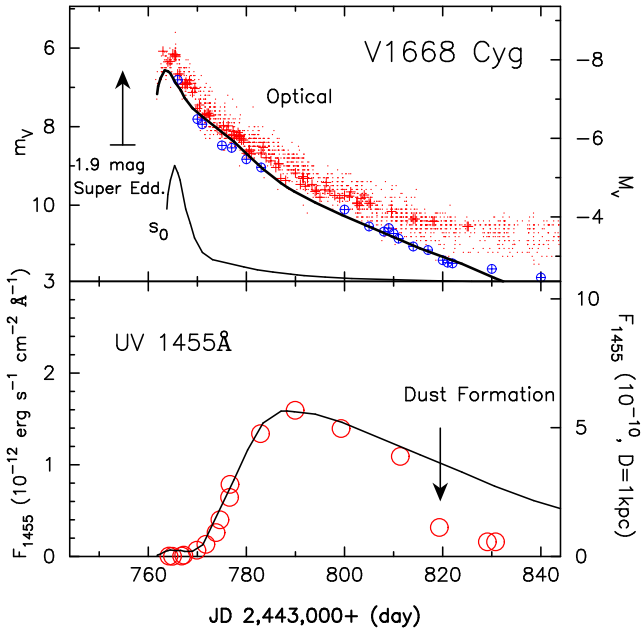}
\caption{
Same as Fig.\ref{v1974cyg2}, but for V1668 Cyg 1978. (a) Upper panel.
Thick solid line: $V$-magnitude from the blackbody photosphere for Model 2.
Thin solid line: the opacity reduction factor $s_0$ 
in the linear scale between 1.0 (at the bottom) and 9.0 (at the peak). 
The WD mass is assumed to be $ 0.95~M_\odot$
with $X=0.45,~X_{\rm CNO}=0.35$, and $Z=0.02$. 
Optical data are taken from AAVSO (dots) and \citet[][crosses]{mal79}.
Data of the Str\"omgren $y$ band magnitude (crosses with a circle) 
are from \citet{gal80}. (b) Lower panel.  
UV data (open circles) are taken from \citet{cas02}. 
The distance of 4.1 kpc is assumed in the upper panel,  
which is obtained from the 1455 \AA~ light-curve fitting in the lower panel.
\label{v1668light}
}
\end{figure}

V1668 Cyg shows a similar but slightly steeper light curve around the 
peak compared with V1974 Cyg. As shown in Figures \ref{v1974cyg2} and
\ref{v1668light}, there is a remarkable difference in 
the 1455 \AA~ fluxes.  In V1668 Cyg the 1455 \AA~ flux remains low in 
several days before rising up, whereas in V1974 Cyg it once decreases 
before the optical maximum and rises again. 
This means that the photospheric temperature in V1668 Cyg remains 
low in a relatively long time around the optical peak. 
This difference led us to find no best fit solutions for Model 1. 
Figure \ref{v1668light} shows our best fit light curve for Model 2, 
where the WD mass  and the chemical composition is assumed to be the 
same as in \citet{hac06}.
This model represents well both the UV and $y$ magnitude light curves.

\subsection{Distance from UV Light-Curve Fitting}

Figure \ref{v1668light} shows that the 1455 \AA~ flux decays quickly 
around JD 2,443,820,  
which can be attributed to the formation of an optically thin dust shell.
\citet{geh80} reported  an excess of the infrared flux that peaks at JD 2,443,815  
whereas no significant drop in the visual magnitude is observed.
Even if the dust shell is optically thin, the grain condensation 
leads to a redistribution of UV flux into infrared.
Considering this effect, which is not included in our 
theory, our $0.95~M_{\odot}$ WD model shows good agreement 
with the 1455 \AA~ observation.

It is difficult to estimate the interstellar absorption in the direction 
of V1668 Cyg because there are only a few star well photometrically observed. 
\citet{hac06} re-examined the distance-reddening law in the direction of 
V1668 Cyg  and estimated the reddening to be $E(B-V) = 0.4$ and obtained the 
distance of 3.6~kpc. 
\citet{sti81} estimated $E(B-V) = 0.4 \pm 0.1$ from 2200 \AA~ feature. 
We adopt $E(B-V) = 0.4$.

The distance is derived to be 4.1 kpc with $E(B-V)=0.4$. 
This value is somewhat larger than 3.6 kpc estimated by 
\citet{hac06} with the normal opacity. In our model, the opacity
reduction factor is still larger than unity ($s_0=1.45$ at the 1455 \AA~ 
peak) as shown in the upper panel of Figure \ref{v1668light}.  Both 
the bolometric and UV fluxes are enhanced compared with those 
in the model of 
\citet{hac06}. Therefore, we obtain a larger distance. 

\citet{slo79} estimated the distance to be 3.3 kpc.  \citet{due80}
obtained a much smaller distance of $d= 2.3$ kpc based on the same
stars. It is very difficult to estimate the  distance because of  
the small number of stars and the very patchy $A_{\rm V}(r)$ relation 
in the direction of the nova \citep[see][]{hac06}. \citet{gal80} derived the 
distance to be 4.4 kpc adopting $M_V=-7.8$ for $E(B-V)=0.3$. However, 
\citet{gal80} preferred a smaller distance of $d \sim 2$ kpc, which was 
derived assuming that the peak luminosity is equal to
the Eddington luminosity because a star with
the super-Eddington luminosity may be dynamically unstable.
This argument cannot be applied to our model, however, 
because the envelope is settled
down into a steady-state even in the super-Eddington phase. 
\citet{sti81} suggested the distance to be 2.2 kpc by equating
the maximum luminosity and the Eddington luminosity.
They also obtained a distance of 3.6 kpc from the relation
between the maximum magnitude and the rate of decline.
They did not take this larger distance because the large acceleration
of matter cannot be expected in the optically thin region. 
However, this is not the case of our optically thick wind. Considering 
these arguments, we conclude that our distance of 4.1 kpc 
is reasonable. 

Using the distance of 4.1 kpc obtained from the 1455 \AA~ light-curve 
fitting, we derive the peak magnitude of $M_{V, \rm max}= -7.73$, i.e.,
super-Eddington by 1.9 mag.

\subsection{Optical Light Curve}

Our theoretical $V$-magnitudes are shown in the upper panel of 
Figure \ref{v1668light}. This curve is placed so as   
to satisfy the distance modulus of 
$(m-M)_{\rm V} = 1.24 +5\log (4.1 {\rm kpc}/10~{\rm pc}) = 14.30$.

There are rich observational data of visual magnitude.
However, we focus on the Str\"omgren $y$ band light curve
\citep[taken from][denoted by the crosses with a circle]{gal80},
which lies along the bottom edge of the visual data,  
because the $y$ filter is designed to avoid strong emission lines
in the nebular phase and reasonably represents the continuum fluxes of novae. 
We regard that our light curve follows the $y$ magnitude because 
our model ignores such line contributions.

The nova enters a coronal phase 53 days
 after the optical maximum \citep{kla80}  
and the spectrum shows strong nebular emission lines \citep{kal86}.
After that, the difference between visual and $y$ magnitudes becomes 
significant as shown in Figure  \ref{v1668light}.

The spectral development of V1668 Cyg can be understood with our model as 
follows.  
The spectrum near the optical maximum was reported to be similar to that of
an intermediate F star  \citep{ort78}, and also be consistent with 
the principal spectrum which is characterized by 
weak hydrogen emission lines and absorption 
lines of neutral and singly ionized metals \citep{kla78}.
In our model, the photospheric temperature is as low as $\log T_{\rm ph}=3.87$,
most of hydrogen is recombined in the region around the photosphere. 
Therefore, these properties are consistent with the observed features.
When the magnitude declined by about 0.8 mag (three days after the maximum), 
\citet{kla80} reported that the nova shows the 
diffuse enhanced spectrum in which strong emission features dominate. 
In our model the photospheric temperature rises to $\log T_{\rm ph}=3.92$ 
at this time and the ionization degree of hydrogen
is quickly increasing with time. 
This is consistent with the appearance of strong H$\beta$  emission.
The nova enters a coronal phase 53 days
after the optical maximum \citep{kla80}.
In our model the temperature rises to $\log T_{\rm ph}=4.57$
at this time and the photon flux is dominated by UV.

The characteristic properties of the light curve are summarizes in Table 1.
The mass of the envelope expelled in the outburst is estimated 
to be $2 \times 10^{-5}M_{\odot}$ \citep{geh80}
and  $5.5 \times 10^{-5}M_{\odot}$
\citep{sti81}. Our model gives an ejecta mass of $5.8 \times 
10^{-5}M_{\odot}$, which is roughly consistent 
with the observational estimates.





\section{V351 PUP (NOVA PUPPIS 1991)} \label{v351pup}

\begin{figure}
\epsscale{1.15}
\plotone{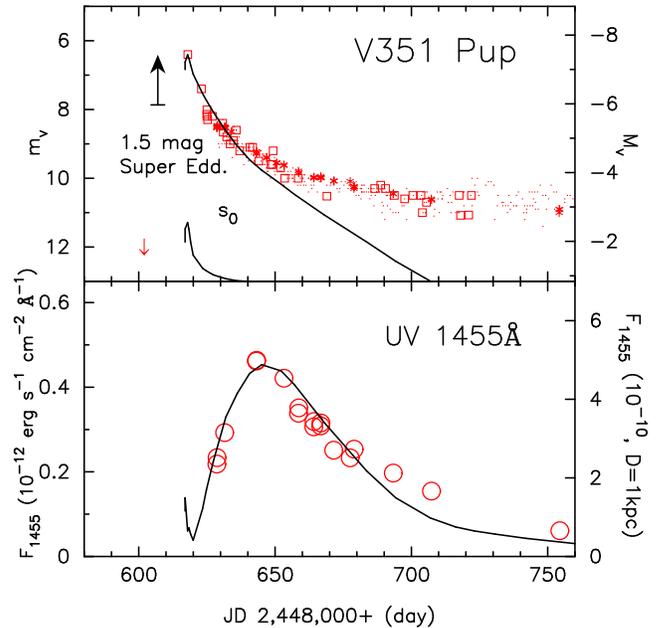}
\caption{
Same as Fig.\ref{v1974cyg2}, but for V351 Pup 1991. (a) Upper panel.
Thick solid line: $V$-magnitude from the blackbody photosphere. 
Thin solid line: the opacity reduction factor $s_0$ 
in the linear scale between 1.0 (at the bottom) and 6.0 (at the peak). 
Optical data are taken from AAVSO (dots), 
{\it IUE} VFES \citet[][asterisks]{cas04a}, and IAUC 5422, 5423, 5427, 
5430, 5437, 5447, 5455, 5493, 5503, and 5527 (squares and arrows). 
(b) Lower panel. Data of UV 1455 \AA~ continuum are taken from \citet{cas02}.
The distance of 2.1 kpc is assumed in the upper panel, 
which is obtained from the 1455 \AA~ light-curve fitting in the lower panel.
\label{v351puplight}
}
\end{figure}

V351 Pup was discovered on 1991 December 27 by \citet{cam92} near maximum.
The light curve shown in Figure \ref{v351puplight} resembles
that of V1668 Cyg in Figure \ref{v1668light}. 
In our modeling we adopt $X=0.35$, $Y=0.23$, $X_{\rm CNO}=0.30$,
$X_{\rm Ne}=0.10$ and $Z=0.02$ after Saizar et al.'s (1996) estimate
from the {\it IUE} spectra, i.e., $X=0.36$, $Y=0.24$,
$X_{\rm CNO}=0.26$, and $X_{\rm Ne}=0.12$.
The reddening is obtained to be $E(B-V)=0.79-0.92$ from emission-line 
ratios \citep{wil94} and  $E(B-V)=0.72\pm0.1$ from 
ratios of recombination lines \citep{sai96}. 
Here we take $E(B-V)=0.72$. 


As optical data are poor around the maximum and no $y$ magnitude data were 
reported, we assume Model 2 as in V1668 Cyg and a slightly larger WD 
mass of $1.0 M_{\odot}$. 
The resultant light curve is shown in Figure \ref{v351puplight} 
which shows good agreement with both the 1455 \AA~ and optical light curves. 
The distance  is estimated to be 2.1 kpc from the UV light-curve fitting.

\section{OS AND (NOVA ANDROMEDAE 1986)} \label{osand}

\begin{figure}
\epsscale{1.15}
\plotone{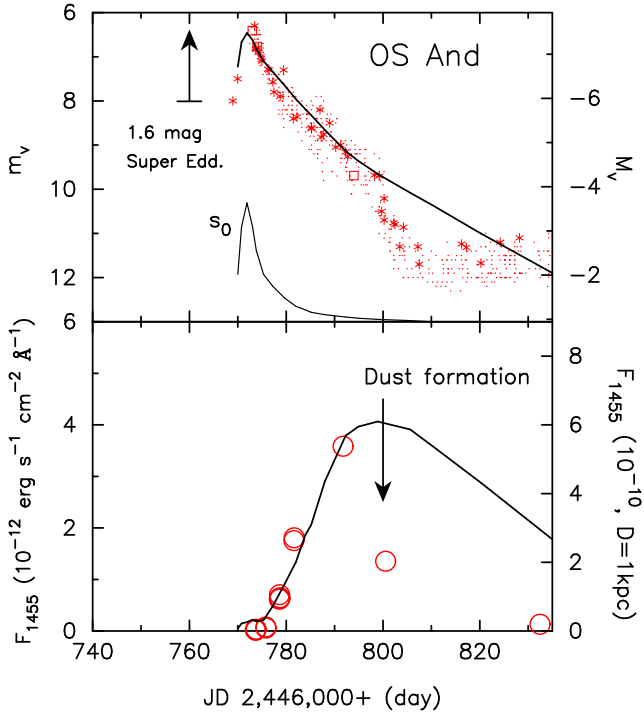}
\caption{
Same as Fig.\ref{v1974cyg2}, but for OS And 1986. (a) Upper panel.
Thick solid line: $V$-magnitude from the blackbody photosphere.
Thin solid line: the opacity reduction factor $s_0$ 
in the linear scale between 1.0 (at the bottom) and 6.0 (at the peak). 
Optical data are taken from AAVSO (dots), \citet[][squares]{kik88},
and IAUC 4281, 4282, 4286, 4293, 4298, 4306, 4342, and 4360 (asterisks). 
(b) Lower panel. The 1455 \AA~ data are taken from \citet{cas02}.
The distance of 4.3 kpc is assumed in the upper panel,  
which is obtained from the fitting in the lower panel.  
\label{osandlight}
}
\end{figure}

OS And was discovered by Suzuki on 1986 December 5  \citep{kos86}. 
The optical light curve in Figure \ref{osandlight} 
shows a 1.5 mag dip that lasts about 30 days owing to dust formation. 
Corresponding to this dip, the 1455 \AA~ light curve  
shows a quick decrease at JD 2,446,800 \citep{cas02}. 
Apart from the deep dip, the optical and 1455 \AA~ light curves resemble
to those of V1668 Cyg and V351 Pup in the first 20 days. 
Moreover, the UV spectrum  is very similar
to that of V351 Pup in the first two weeks \citep{son92}. 


With such a strongly absorbed UV light curve, it is 
difficult to search for a best fit model.  Therefore, we assume 
the same WD mass and the same $s$ function as those in V351 Pup. 
The chemical composition is assumed to be the same as in V1668 Cyg, 
regarding OS And as a CO nova.  Note that the difference in 
the composition does not make large difference as shown in the two light 
curves of  V1668 Cyg and V351 Pup. 
The blackbody light curve in Figure \ref{osandlight} shows
good agreement with both the visual and UV data in the first 20 days. 

The distance is estimated to be 4.3 kpc
from the 1455 \AA~ light curve fitting in 
the lower panel for a reddening of  $E(B-V)=0.25$ \citep{sch97}. 
Our value of 4.3 kpc is roughly consistent with $5.1 \pm 1.5$~kpc  
obtained from a comparison of UV fluxes between \object{OS And} and
\object{Nova LMC 1992} \citep{sch97}.



\section{DISCUSSION AND SUMMARY} \label{discussion}

Table 1 summarizes the model parameters and our main results, i.e.,
from top to bottom row,  (1) object name, (2) outburst year,
(3) adopted function of the opacity reduction factor,  
(4) WD mass, (5)--(9) adopted chemical composition, 
(10) reddening, (11) opacity reduction factor $s_0$ at the optical maximum,
(12) distance estimated from the peak of the 1455 \AA~light-curve, 
(13) duration of the 1455 \AA~ outburst defined by the full width
at the half maximum (FWHM), 
(14) peak bolometric luminosity, 
(15) absolute and (16) apparent $V$-magnitudes 
corresponding to the peak luminosity, 
(17) excess of the super-Eddington in $V$-magnitude, i.e., 
the difference between the peak magnitude of
our model ($M_{V, \rm max}$) 
and the peak magnitude of a light curve model
with the normal opacity ($s\equiv 1$)
for the same WD mass and the same envelope composition,  
(18) duration of the super-Eddington phase,
(19) time in which  $M_V$ drops by 3 magnitude
from the peak in our theoretical model, 
(20) mass ejected during the period from the first point
of each light curve until the wind stops.   


\begin{figure}
\epsscale{1.15}
\plotone{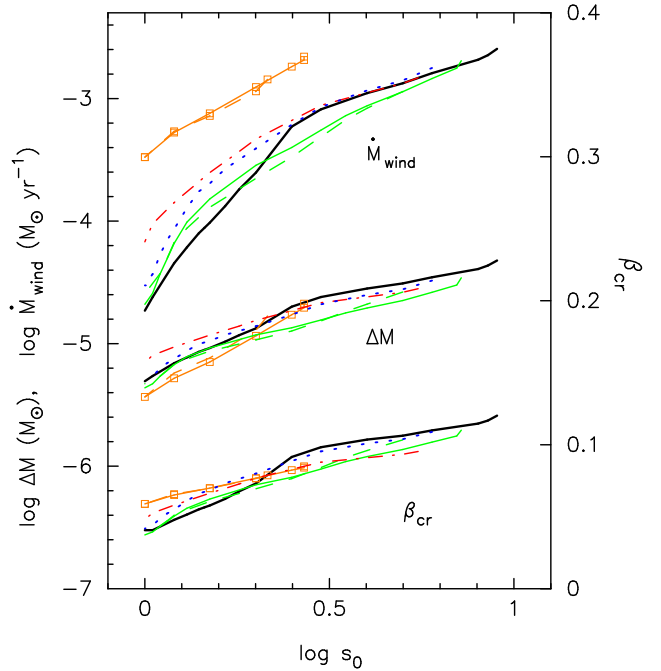}
\caption{
The wind mass-loss rate ($\dot M_{\rm wind}$),
envelope mass ($\Delta M$), and the ratio of the gas pressure to the total
pressure at the critical point of steady-state wind ($\beta_{\rm cr}$)
are plotted against the opacity reduction factor, $s_0$,
in the super Eddington phase. The right-end point of each curve corresponds
to the optical peak while the left-end point corresponds to $s_0=1$,
i.e., the epoch when the super Eddington phase ends.
Time goes on from right to left.  
{\it Dashed/Solid line with squares}: Model 1/Model 2
of V693 CrA, respectively.
Both of them are almost overlapped in all three cases of
$\dot M_{\rm wind}$, $\Delta M$, and $\beta_{\rm cr}$.
{\it Dashed/Thin solid line}: Model 1/Model 2 of V1974 Cyg. 
{\it Thick solid line}: V1668 Cyg.  
{\it Dash-dotted line}: V351 Pup. 
{\it Dotted line}: OS And.
\label{superedpara}
}
\end{figure}

\subsection{Opacity Reduction Factor}
 
There are no time-dependent calculations of nova outbursts 
in which porous instabilities widely develop.  So, it is difficult
for us to estimate how much the opacity is reduced in the nova envelopes.
Here, we have simply assumed $s$ to be a function of the 
temperature and time.  Then, we have determined $s_0$ so as to reproduce 
the observed light curve of each object. 

The referee kindly pointed out that the opacity reduction factor
would be expressed as a function of the current state of the envelope 
which is independent of time.  After a porous instability widely develops
in the envelope, it may settle into an equilibrium state.
If our opacity reduction factor represents such an
envelope state, it may be a function of a small number of physical 
parameters/variables that represent the current state of the envelope.

In order to search for such parameters/variables, we plot
the wind mass-loss rate ($\dot M_{\rm wind}$),
envelope mass ($\Delta M$), and ratio of the gas pressure
to the total pressure at the critical point of each steady wind solution
($\beta_{\rm cr}$) against $s_0$ in Figure \ref{superedpara}. 
We plot seven light-curve models for five objects in Table 1. 
These seven models are different from each other in their characteristic 
properties such as the WD mass, chemical composition, and opacity model. 

First of all, we can easily see that the difference between Model 1 
and Model 2 hardly makes large difference on these three variables;
the curves for Model 1 and Model 2 are very close to each other
in V693 CrA and V1974 Cyg. 

Next, we see that the two models of V693 CrA are separated from the others
in the wind mass-loss rate.
This is because the WD of V693 CrA is as massive as
1.3 $M_{\odot}$ whereas the other WDs are $0.95-1.05 ~M_{\odot}$. 
The wind mass-loss rate is much larger on a massive WD
than on a less massive one if we compare them at the same envelope mass,
as shown in our previous work \citep{kat94h}.
Therefore, the difference in the wind mass-loss rates of V693 CrA in Figure
\ref{superedpara} can be attributed to the difference in the WD masses.
So we may conclude that the mass-loss rate itself is not a main factor
that determines the opacity reduction factor $s_0$. 

In the envelope mass, all the curves are almost similar to each other.
This suggests that $s$ is closely related to the envelope mass.
In other words, the porous instability involves at least a wide area
of the envelope, and $s_0$ decreases as the envelope mass decreases
with time due to wind mass-loss.

In the ratio of the gas pressure to the total pressure
at the critical point of steady-state wind solutions,
where the wind is accelerated \citep{kat94h}, 
we see that all the curves are also very similar to each other. 
The ratio $\beta$ is almost constant in a region below the critical point.
Therefore, $\beta_{\rm cr}$ represents a mean value of $\beta$ in the 
region where the opacity is reduced. 

We have checked other physical variables such as the luminosity,
photospheric radius, temperature, and radius at the critical point, 
but we found that all these variables are largely scattered
from each other, i.e., not bunched like in $\Delta M$
and $\beta_{\rm cr}$ of Figure \ref{superedpara}.

We may conclude that the envelope mass and $\beta_{\rm cr}$ are
the key parameters that represent the current state of envelopes
with a porous structure.  Probably, $\beta_{\rm cr}$ should be 
closely linked with the property that the porous instability is
a kind of radiation instability and the envelope mass may be
related with the property that the porous instability involves
a large part of the envelope, although we do still not know
the accurate condition for the porous instability.

\subsection{White Dwarf Mass and Chemical Composition}

Both the WD mass and chemical composition are fixed in the present work.
If we choose a different set of these parameters,
we have different time scales of theoretical light curves
not only in the very early phases but also in the later phases of
the outbursts including the epochs when the wind mass-loss stops
and hydrogen burning ends.  For V1974 Cyg and V1668 Cyg,
multiwavelength observations are available until the
very late phase of the outburst and their WD masses are determined 
consistently with these observations
\citep[see, e.g.,][for V1974 Cyg and V1668 Cyg]{hac05k,hac06}. 
Therefore, we adopt their estimates.

The decline rates of nova light-curve depend strongly on the WD mass 
and weakly on $X$, very weakly on $X_{\rm CNO}$, but hardly on 
$X_{\rm Ne}$. Therefore,  the largest ambiguity in the estimation
of WD masses comes from the accuracy of hydrogen content $X$.
The dependency of the WD mass on $X$ is roughly estimated as  
\begin{equation}
M_{\rm WD}(X) \approx M_{\rm WD}(0.55) + 0.5 (X-0.55),
\label{ne_wd_mass_x_depencency}
\end{equation}
when $0.35 \le X \le 0.65$ and $0.03 \le X_{\rm CNO} \le 0.35$, using 
the ``universal decline law'' of nova light curves \citep{hac06} that
nova light curves are almost homologous except for the very early
phase (i.e., the super-Eddington phase discussed here).   
Here, $M_{\rm WD}(0.55)$ means the WD mass estimated for $X=0.55$
and the original value of $X=0.55$ can be replaced with any other
value, for example, $X=0.35$.

For the other three objects, i.e., V693 CrA, V351 Pup, and OS And,
there is no observational data in the late phase, i.e., when the wind
stopped and when the hydrogen burning ended.  Then we determined
the WD mass only from the UV light curve fitting with a fixed chemical
composition as shown in Figure \ref{v693craUV}.
If we adopt a different set of the chemical composition,
we have a slightly different WD mass as can be estimated from
equation (\ref{ne_wd_mass_x_depencency}) or Figure \ref{UVwidth}.

\subsection{Distance}

Nova distances are determined from the comparison between observed UV
fluxes and calculated fluxes at the UV peak.
In case of V693 CrA and V351 Pup,
the super-Eddington phase ended, i.e., $s_0=1$, at the UV peak
as shown in Figures \ref{v693light} and \ref{v351puplight}.
So, the distance can be determined independently of the reduced factor
of $s$.  In case of V1974 Cyg, the super-Eddington phase still continues
at the UV peak.  The distances derived for Model 1 and Model 2 
are the same within the accuracy of two digits as shown in Table 1, 
although these values themselves are somewhat larger than the values
derived without the super-Eddington phase ($s \equiv 1$) as already
mentioned in \S 5.1.  From these results, it may be concluded
that the opacity reduction factor does not affect so much
the distance estimate. 
This is because, in all the five objects, their super-Eddington phases
had already or almost ended ($s_0 \cong 1.0$) at the UV peak.


\subsection{Summary}

Our main results are summarized as follows;

1. We present light-curve models of the super-Eddington phases for
five {\it IUE} classical novae based on the optically thick wind theory,
with an assumption that the opacity is reduced in a porous 
envelope \citep{sha02}.  Our models reasonably reproduce the optical
and  1455 \AA~ light curves. 

2. The duration of the 1455 \AA~ light curve is a useful indicator of the WD
mass, especially when the chemical composition is known.

3. The distance is derived from the comparison between the observed peak 
value of the 1455 \AA~ flux and the corresponding calculated value 
unless a dust shell absorbs the UV flux at its maximum.  The derived
distances of the five IUE novae are consistent with the previous 
estimates.




\acknowledgments

We wish to thank the American Association
of Variable Star Observers (AAVSO) for the visual data of 
V693 CrA, V1974 Cyg, V1668 Cyg,
V351 Pup, and OS And, and also the Variable Star Observing League
of Japan (VSOLJ) of \object{V693 CrA}.
We also thank A. Cassatella for providing machine readable 1455 \AA~
data of the five novae.  
We thank the anonymous referee for useful and valuable comments that
improved the manuscript.
This research has been supported in part by the Grant-in-Aid for
Scientific Research (16540211, 16540219)
of the Japan Society for the Promotion of Science.

\end{document}